\documentclass[aip,rsi,amsmath,amssymb,reprint,floatfix]{revtex4-1}

\usepackage{graphicx}
\usepackage{dcolumn}
\usepackage{bm}

\usepackage{siunitx}

\usepackage{url}
\usepackage[nooneline,raggedright]{subfigure}

\begin{document}

\title[Electrostatic deflector studies using small-scale prototype electrodes]{Electrostatic deflector studies using small-scale prototype electrodes}

\author{K. Grigoryev}
  \email{k.grigoryev@fz-juelich.de}
  \affiliation{Institute for Nuclear Physics, Forschungszentrum J\"ulich, 52425, J\"ulich, Germany}

\author{F. Rathmann}
  \affiliation{Institute for Nuclear Physics, Forschungszentrum J\"ulich, 52425, J\"ulich, Germany}

\author{A. Stahl}
  \affiliation{III Physics Institute B, RWTH Aachen University, 52074, Aachen, Germany}

\author{H. Str\"oher}
  \affiliation{Institute for Nuclear Physics, Forschungszentrum J\"ulich, 52425, J\"ulich, Germany}

\date{\today}

\begin{abstract}
The search for electric dipole moments of particles in storage rings requires the development of dedicated deflector elements with electrostatic fields. The JEDI prototype-ring design consist of more than 50 electric deflectors of \SI{1}{\meter} length with \SI{60}{\milli \meter} spacing between the plates with electric field gradients of \SI{10}{\mega \volt \per \meter}. This paper presents studies of scaled-down uncoated prototype electrodes of \SI{10}{\milli \meter} in diameter made from stainless steel. The electric field at distances from $\SI{1}{\milli \meter}$ down to $\SI{0.05}{\milli \meter}$ increased from 15 to $\SI{90}{\mega \volt \per \meter}$. In future investigations we will also study different materials and coatings at similar distance between the electrodes. Preparations are underway to study also large deflector elements.
\end{abstract}

\maketitle

\section{Introduction}
\label{sec:introduction}

The JEDI collaboration\footnote{J{\"u}lich Electric Dipole moment Investigations, http://collaborations.fz-juelich.de/ikp/jedi} is searching for permanent electric dipole moments (EDM) of charged particles, such as protons\,\cite{doi:10.1063/1.4967465} and deuterons\,\cite{Rathmann:2011zz}. One of the technical challenges is the development of electric bending elements that provide high electric fields. A purely electrostatic EDM ring of \SI{30}{\meter} radius, for instance, requires electric fields of about \SI{17}{\mega \volt \per \meter}\,\cite{Rathmann:2014gfa}. The present limit for the electric field of bending elements at accelerators is below \SI{10}{\mega \volt \per \meter}\,\cite{reiser2008theory}. The electrostatic separators at CESR\footnote{Cornell Electron-positron Storage Ring, https://www.classe.cornell.edu/public/lab-info/cesr.html}\,\cite{Welch:1999qs} and Fermilab\footnote{Fermi National Accelerator Laboratory, http://www.fnal.gov} Tevatron\,\cite{lebedev2014accelerator,Shiltsev:2005xa}, and the CERN septa\,\cite{Borburgh:2003ke} are routinely operated at smaller electric fields.

In order to study different electrode materials and coatings, the investigations described here made use of scaled-down prototypes and a dedicated UHV test stand installed inside a clean room at RWTH Aachen University. The operation in the laboratory with respect to radiation protection was simplified, because the applied voltages were always below $\SI{30}{\kilo \volt}$. Nevertheless, by scaling down the applied voltage and by reducing at the same time the spacing between the electrodes, large electric fields could be obtained.

In Sec.\,\ref{sec:experimental-setup}, the experimental setup is described. General considerations of the deflector development are given in Sec.\,\ref{sec:shape-size}. The electrical scheme using a high-voltage power supply is discussed in Sec.\,\ref{sec:electric-scheme}, and the set up of the vacuum system inside a clean room is described in Sec.\,\ref{sec:vacuum-system}. The electrodes are presented in Sec.\,\ref{sec:deflector-prototypes}, and the measurements  in Sec.\,\ref{sec:measurements}. The results are summarized in Sec.\,\ref{sec:summary}.

\section{Experimental setup}
\label{sec:experimental-setup}

\subsection{Small-scale prototype electrodes}
\label{sec:shape-size}

Initial investigations about the shape of electrostatic deflectors were based on existing elements used at the Fermilab Tevatron\,\cite{prokofiev2005}. The plates of the Tevatron electrostatic separators were designed to provide a field of \SI{6}{\mega \volt \per \meter} at distances of $40 - \SI{60}{\milli \meter}$ with a length of about \SI{2.5}{\meter}. The transverse profile of the Tevatron separators represents a Rogowski profile\,\cite{4113920}. For a specific electrode configuration (\textit{i.e.,} plate separation and height) the surface contour of the electrodes is designed to follow the equipotential lines. Such a profile ensures a high homogeneity of the electric field in the flat region between the deflector plates, and, according to Refs.\,\cite{Rogowski:1923,Rogowski:1926}, a discharge will occur \textit{outside} of that region.

In order to simplify the mechanical production of prototype elements for the test setup at RWTH Aachen University, all elements were manufactured with round corners rather than with Rogowski profiles. The smallest elements consist of half-spheres with a radius of $R = \SI{10}{\milli \meter}$. Small test samples also served to minimize weight which eliminated the need for a sophisticated support structure.

\subsection{Electrical scheme}
\label{sec:electric-scheme}

\begin{figure}[tb]
	\begin{minipage}{0.24\textwidth}
		\begin{flushleft}
			\includegraphics[height=0.26\textheight]{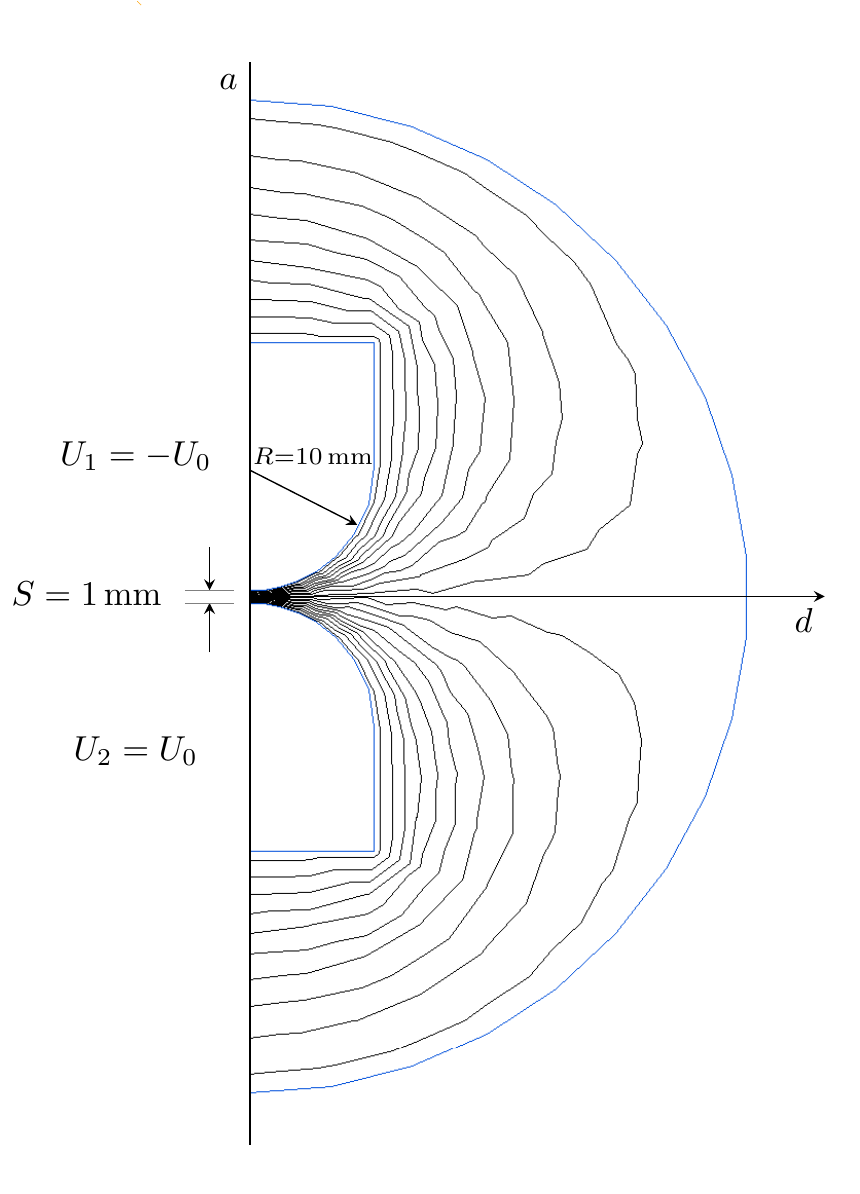}
			(a) Electrodes at $U_1 = -U_0$\\and $U_2 = U_0$.
			\label{fig:electrodes-symmetric}
		\end{flushleft}
	\end{minipage}\hfill
	\begin{minipage}{0.24\textwidth}
		\begin{flushleft}
			\includegraphics[height=0.26\textheight]{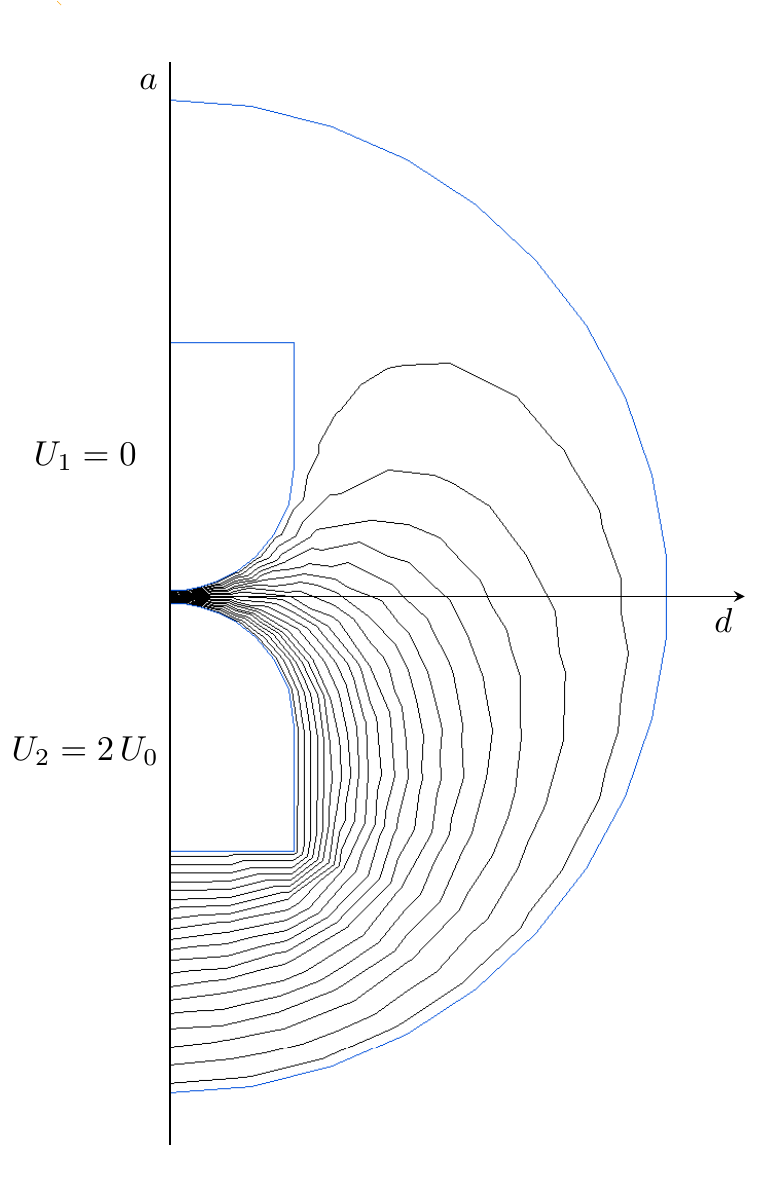}
			(b) Electrodes at $U_1=0$\\and $U_2 = 2~U_0$.
			\label{fig:electrodes-one-grounded}
		\end{flushleft}
	\end{minipage}
	\begin{flushleft}
		\includegraphics[width=1\columnwidth]{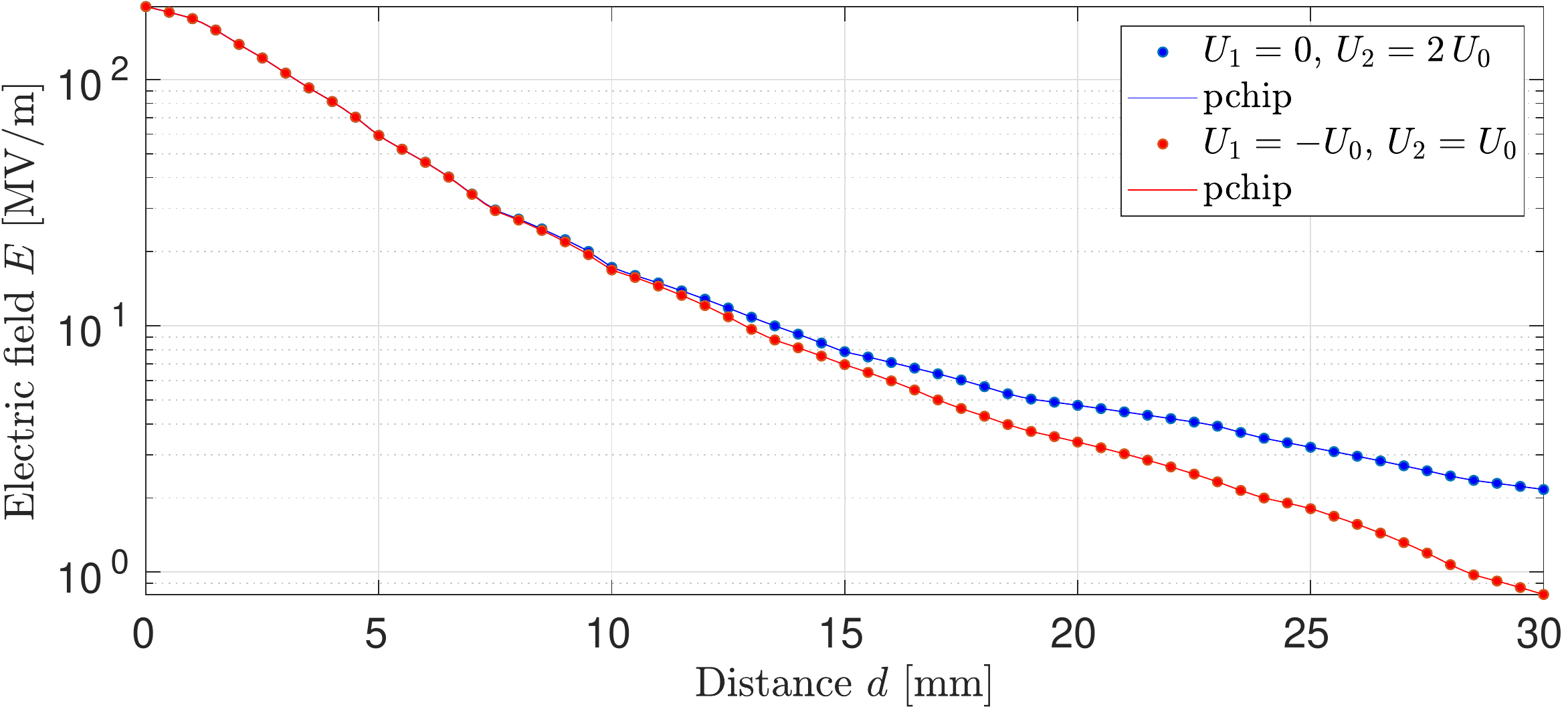}
		\label{fig:electric-field-graph}
		(c) Electric field $E$ as function of distance $d$ for the geomet- ries shown in (a) and (b), with a piecewise cubic hermite interpolating polynomial (pchip).
		\caption{Equipotential lines using QuickField\cite{Note4} for the configurations shown in panels (a) and (b) with $R=\SI{10}{mm}$, $S = \SI{1}{mm}$, and $U_0 = \SI{10}{\kilo \volt}$. The electrodes are rotationally symmetric around the axis $a$. The electric field $E$ along $d$ for both cases is depicted in panel (c).}
		\label{fig:electrodes-symmetric-asymmetric}
	\end{flushleft}
\end{figure}

Numerical simulations of the electric field $E$, performed with the QuickField FEA software\footnote{QuickField simulation software, Tera Analysis Ltd, https://quickfield.com}, showed that for our studies there is essentially no difference between using the same but opposite potentials $U_0$ between the electrodes or having one of the electrodes powered with twice the voltage $2~U_0$, while the other one is grounded (see Fig.\,\ref{fig:electrodes-symmetric-asymmetric}).

The option with one grounded electrode (see Fig.\,\ref{fig:electrodes-symmetric-asymmetric}(b)) is more attractive, because the measurements can then be performed with a single high-voltage power supply. In addition, common ground for every device minimizes the measurement noise and makes the  dark current detection with a picoammeter\footnote{Keithley picoammeter 6485, https://www.tektronix.com} more reliable. 

\begin{figure}[tb]
	\centering
	\includegraphics[width=1\columnwidth]{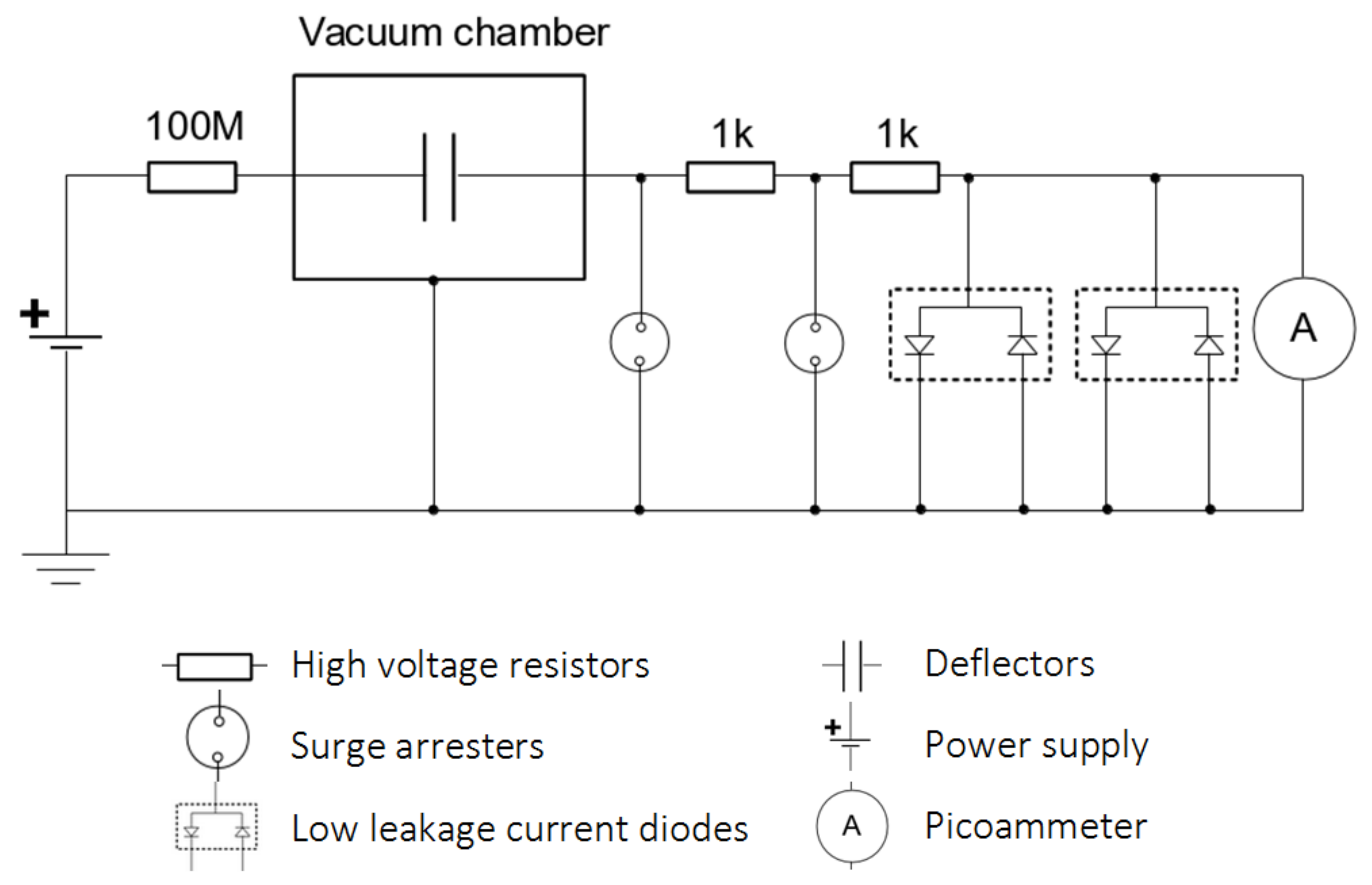}
	\caption{Electrical schematic of the test apparatus used for the measurements. A~\SI{100}{\mega \ohm} resistor, two gas discharge boxes act as surge arresters\cite{Note6} and two low-leakage diodes\cite{Note7} are used to protect the power supply and the picoammeter during voltage breakdown in an electric discharge.}
	\label{fig:electric-scheme}
\end{figure}

The electrical circuit, shown in Fig.\,\ref{fig:electric-scheme}, consists of two discharge protection elements. The \SI{100}{\mega \ohm} resistor serves to limit the current to ground during HV breakdown. The gas discharge boxes\footnote{EPCOS AG, type EPCOS EC90X and EC600X, https://tdk-electronics.tdk.com} and the low-leakage diodes\footnote{Diodes Incorporated, https://www.diodes.com, type BAV199} protect the picoammeter from high currents during a discharge.

To ensure safe operation, the high-precision high-voltage power supply\footnote{Heinzinger PNC 30000, https://www.heinzinger.com/products/high-voltage/universal-high-voltage-power-supplies} was equipped with a rapid discharge circuit. A fast discharging capacitor within the power supply reduces the voltage to less than 1\% of the applied value in less than \SI{1}{\second}. The measured voltage ripple of the power supply was below the specified value of \num{e-4} at \SI{30}{\kilo \volt} and was stabilized to better than 0.05\% over a time interval of \SI{8}{\hour}.

\subsection{Clean room and vacuum system}
\label{sec:vacuum-system}

\begin{figure}[tb]
	\centering
	\includegraphics[width=1\columnwidth]{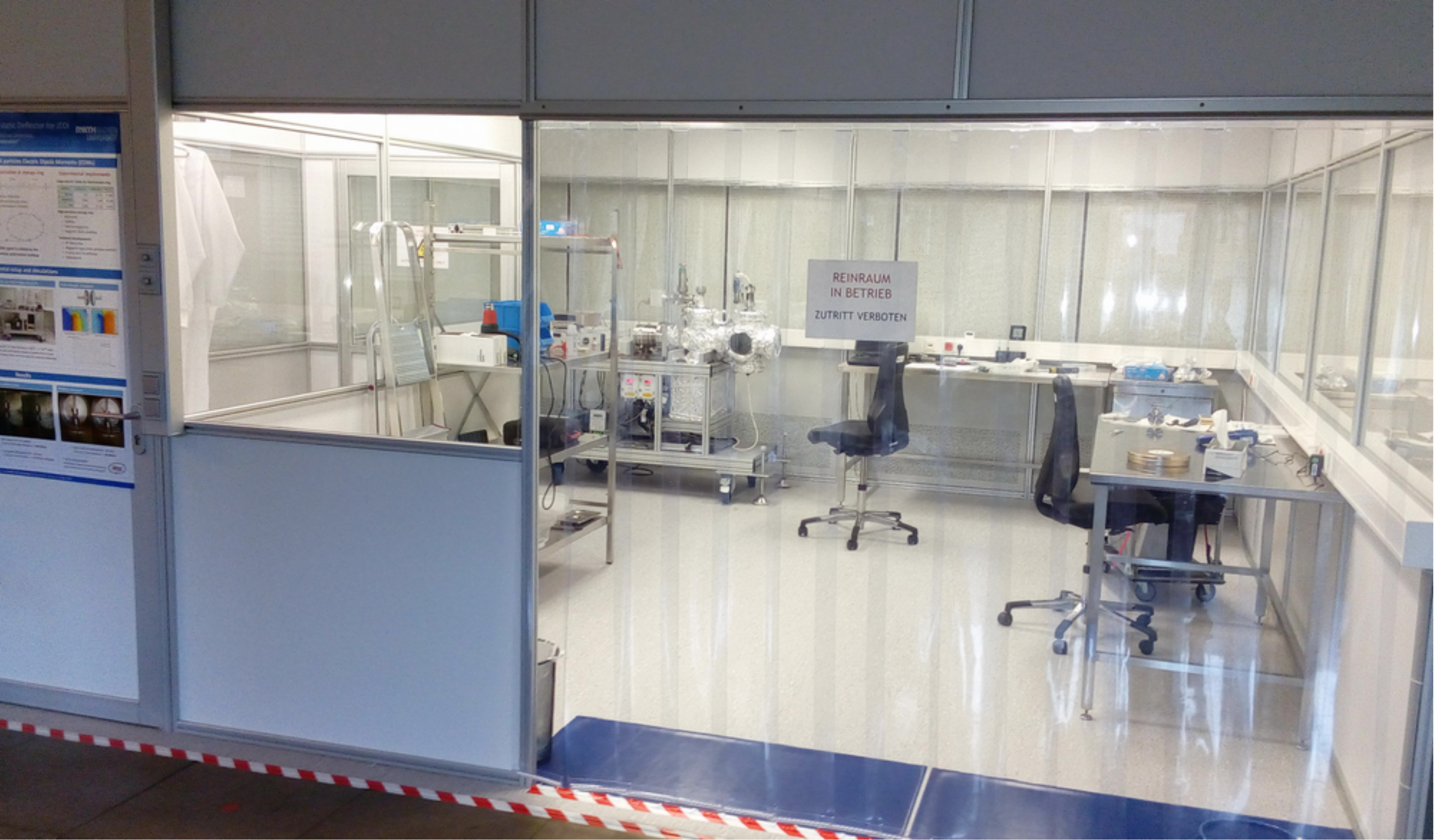}
	\caption{Class ISO7\cite{Note9} clean room at RWTH Aachen University, housing the experimental setup for electrostatic tests.}
	\label{fig01b}
\end{figure}

To perform the test measurements, a \SI{25}{\square \meter} class ISO7\footnote{A class ISO7 clean room allows inside \SI{1}{\cubic \meter} of air, a maximum of \SI{e7}{particles} of size $>\SI{0.1}{\micro \meter}$, and not more than \SI{352000}{particles} of size $>\SI{0.5}{\micro \meter}$.} clean room was installed in the experimental hall at RWTH Aachen University with a gateway and a strip curtain for rolling the test apparatus (see Fig.\,\ref{fig01b}). The clean area inside was sufficiently large to place a few tables besides the test stand for the prototypes.

\begin{figure}[tb]
	\begin{center}
		\includegraphics[width=1\columnwidth]{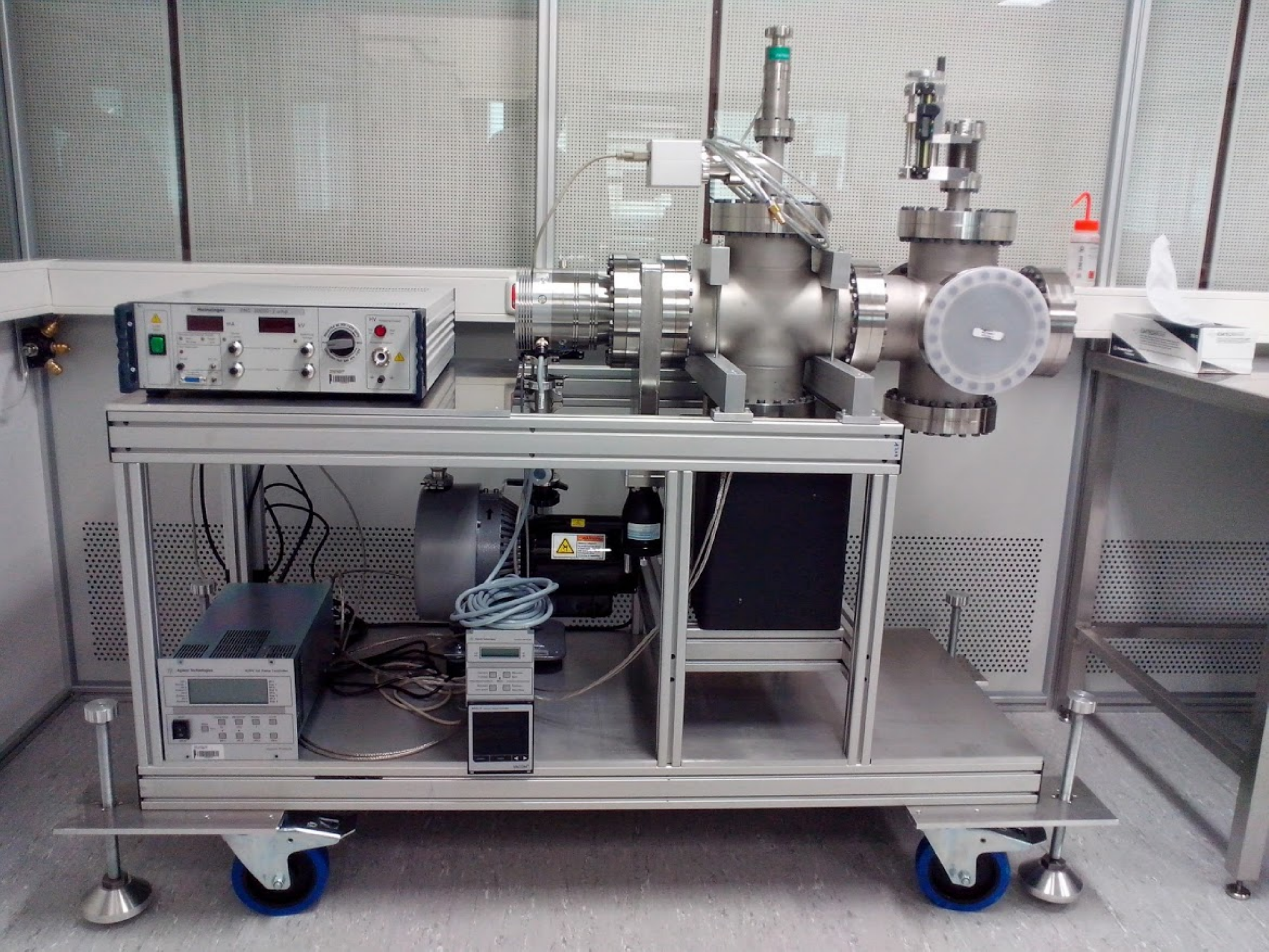}
	\end{center}
	\begin{flushleft}
		(a) Photograph of the experimental test bench.
		\label{fig:test-bench-foto}
	\end{flushleft}
	\begin{center}
		\includegraphics[width=0.85\columnwidth]{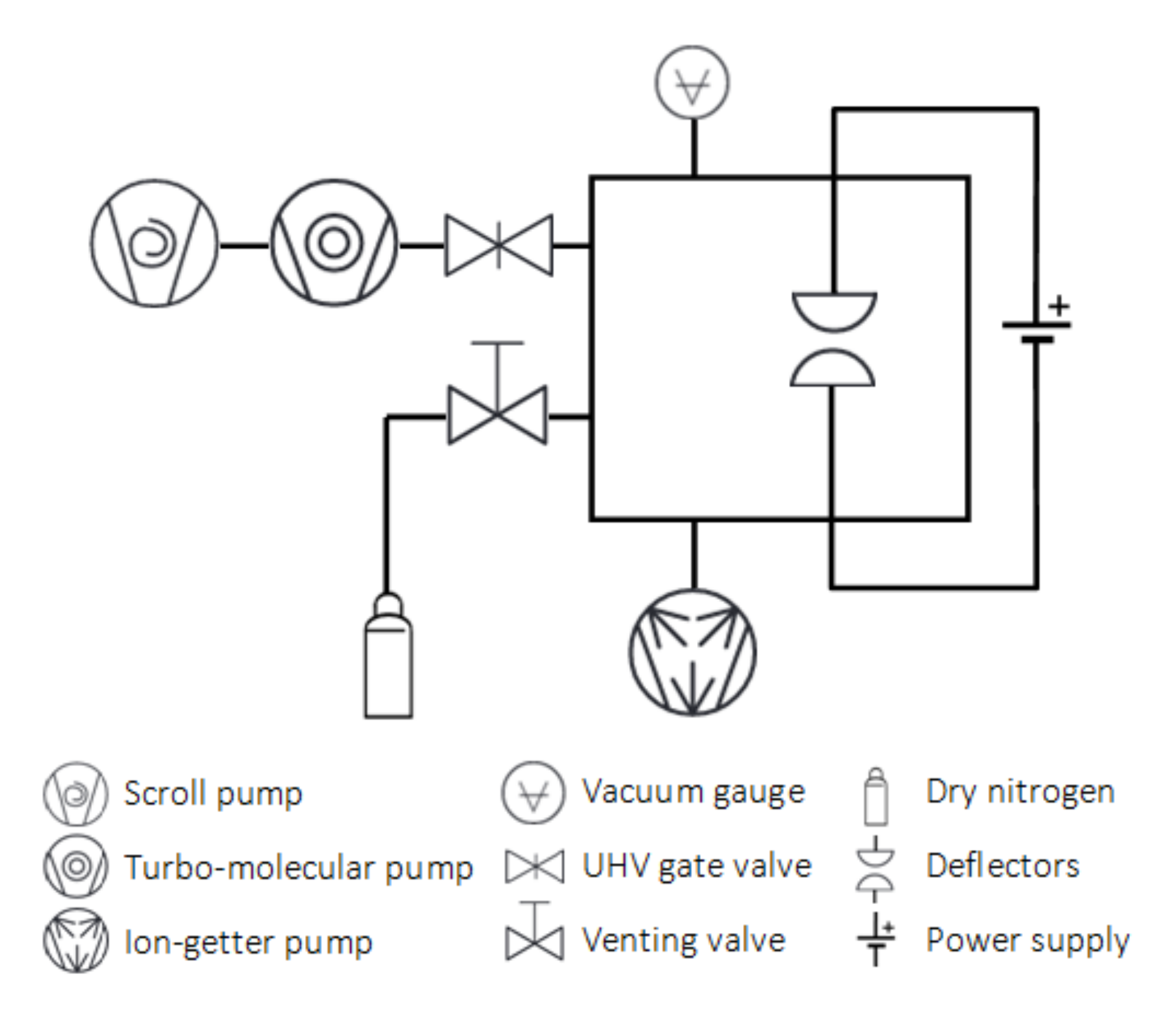}
	\end{center}
	\begin{flushleft}
	    (b) Schematic of the vacuum system.
	    \label{fig:test-bench-scheme}
		\\
		\caption{Experimental test bench in the clean room to study small scale prototype electrodes at RWTH Aachen University.}
		\label{fig:test-bench}
	\end{flushleft}
\end{figure}

A dust-free vacuum system was designed and built using UHV components, mounted on a movable support for easy access and flexibility during the tests measurement (see Fig.\,\ref{fig:test-bench}(a)). An oil-free turbo-molecular pump\footnote{Agilent TwisTorr 304, https://www.agilent.com} with \SI{300}{\ell \per \second} pumping speed and air cooling, backed by a dry scroll pump\footnote{Agilent TriScroll~300, https://www.agilent.com} with \SI{15}{\cubic \meter \per \hour} pumping speed allowed us to reach good vacuum conditions within a few minutes. Simultaneous heating of the chamber to the maximum operating temperature of the turbo-pump~(\SI{80}{\degreeCelsius}) removed water from the stainless-steel walls of the vacuum chamber and brought the pressure down to about~\SI{e-9}{\milli \bar}.

A \SI{300}{\ell \per \second} ion-getter pump\footnote{Agilent Vaclon Plus~300, https://www.agilent.com}, installed directly on the vacuum chamber, equipped with its own heater, was used to activate the ion-getter pump at the same time when the vacuum chamber was baked out. After activation of the ion-getter pump, the vacuum chamber was isolated from the scroll and turbo-molecular pumps using a UHV gate-valve\footnote{Vacom 5GVM-160CF-MV-S, copper sealed ultra-high vacuum hand gate-valve, https://www.vacom.de} (see Fig.\,\ref{fig:test-bench}(b)), and the pressure reached about \SI{e-11}{\milli \bar}. During the tests the scroll and turbo-molecular pumps were turned off to minimize vibrations. The pressure in the vacuum chamber, measured directly by the ion-getter pump, was typically of the order of \SI{e-10}{\milli \bar}.

\section{Test electrodes}
\label{sec:deflector-prototypes}

\begin{figure}[tb]
	\centering
	\includegraphics[width=0.90\columnwidth]{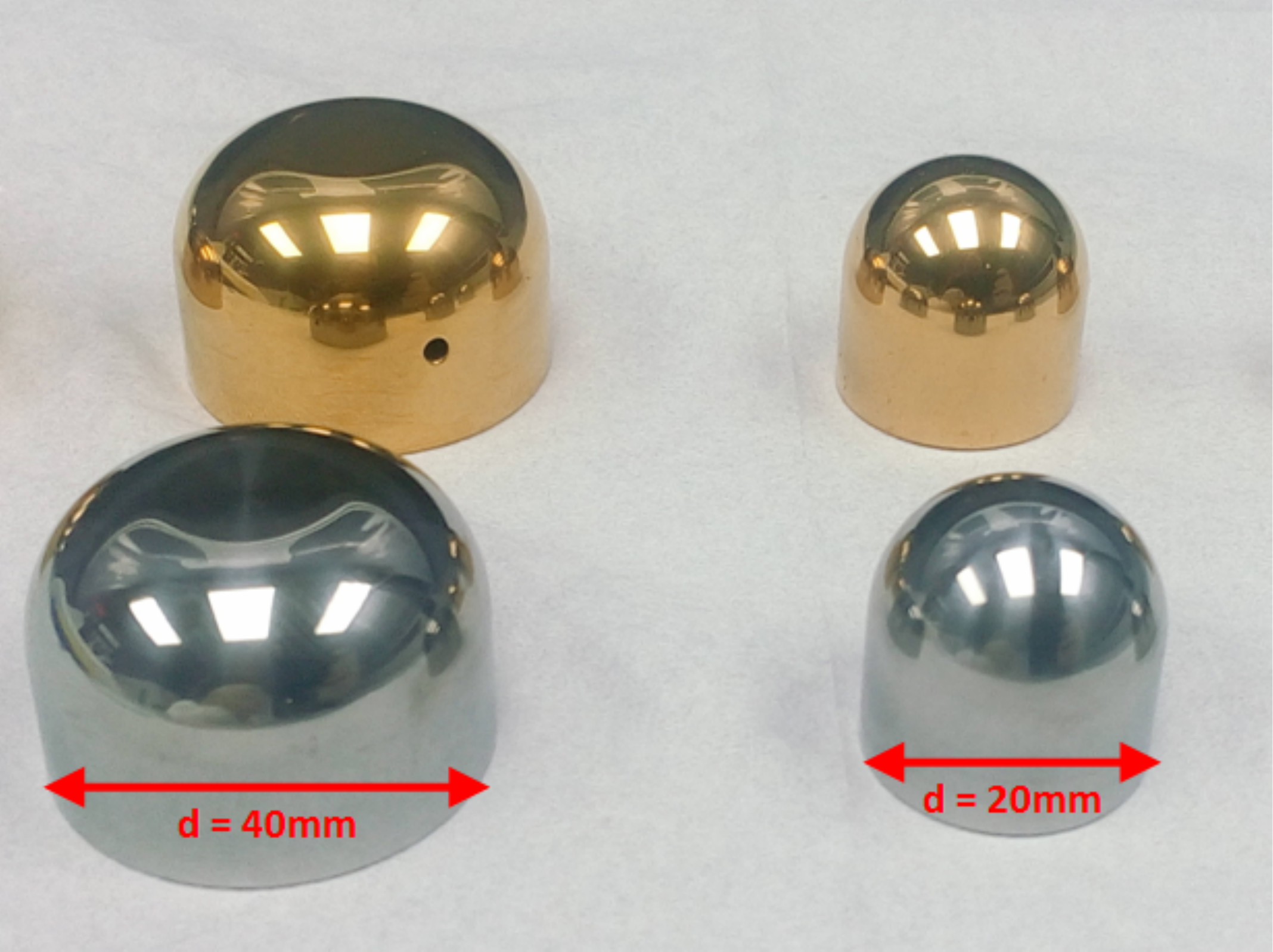}
	\caption{Stainless-steel electrode prototypes. The uncoated elements are shown in the front, the TiN coated in the back.}
	\label{fig:SS-deflector-prototypes}
\end{figure}

The electrodes for the tests were made of stainless steel in two different sizes. The small ones are half-spheres of radius $R = \SI{10}{\milli \meter}$. The large electrodes additionally possess a flat central region of $\SI{20}{\milli \meter}$ diameter. Based on the experience reported in\,\cite{FURUTA200533,doi:10.1116/1.4916574}, for further investigations a set of stainless-steel samples prepared in the same way was coated by TiN (see Fig.\,\ref{fig:SS-deflector-prototypes}). The results of the measurements using stainless steel and aluminum coated with TiN will be described in a forthcoming publication.

The test electrodes were produced and mechanically polished in the RWTH Aachen workshop. The average roughness of the surface was smaller than $\SI{0.10}{\micro \meter}$ with a maximum nonuniformity of $\SI{1.17}{\micro \meter}$. Prior to installation into the vacuum chamber, all parts were cleaned in an ultrasonic propanol bath.

For precise positioning, each measurement started by mechanically setting the distance between the electrodes to zero. This was accomplished by applying a small voltage and observation of a large current at the picoammeter when the electrodes touched each other. From there, one of the electrodes was moved to the measurement position using a manual UHV-compatible linear drive\footnote{Compact linear drive CLSM38-50-H-DLA from UHV Design, http://www.uhvdesign.com} with \SI{0.01}{\milli \meter} positioning accuracy.

The electric field between the two half-spherical electrodes can be written as\,\cite{Russell1927}
\begin{equation}\label{eq:field-enhancement}
E_\text{max} = \frac{U}{S} \cdot F~,
\end{equation}
where $F$ denotes the so-called field enhancement factor, $U$ the voltage, and $S$ the spacing between the electrodes. The field enhancement factor $F$ can be calculated for known shapes. For half-spherical electrodes with radius of curvature $R$\,\cite{Dean1913},
\begin{equation}\label{eq:field-enhancement-factor} 
F =  \frac{1}{4}\left[ 1 + \frac{\displaystyle S}{\displaystyle R} + \sqrt{ \left(1 + \frac{S}{R}\right)^2 + 8}\right]~,
\end{equation}
where $S$ denotes the spacing between the two spheres, so that the distance between the centers of the half-spheres is given by 
\begin{equation}\label{eq3}
D = S + 2 R~.
\end{equation}

\begin{figure}[tb]
	\centering
	\includegraphics[width=1\columnwidth]{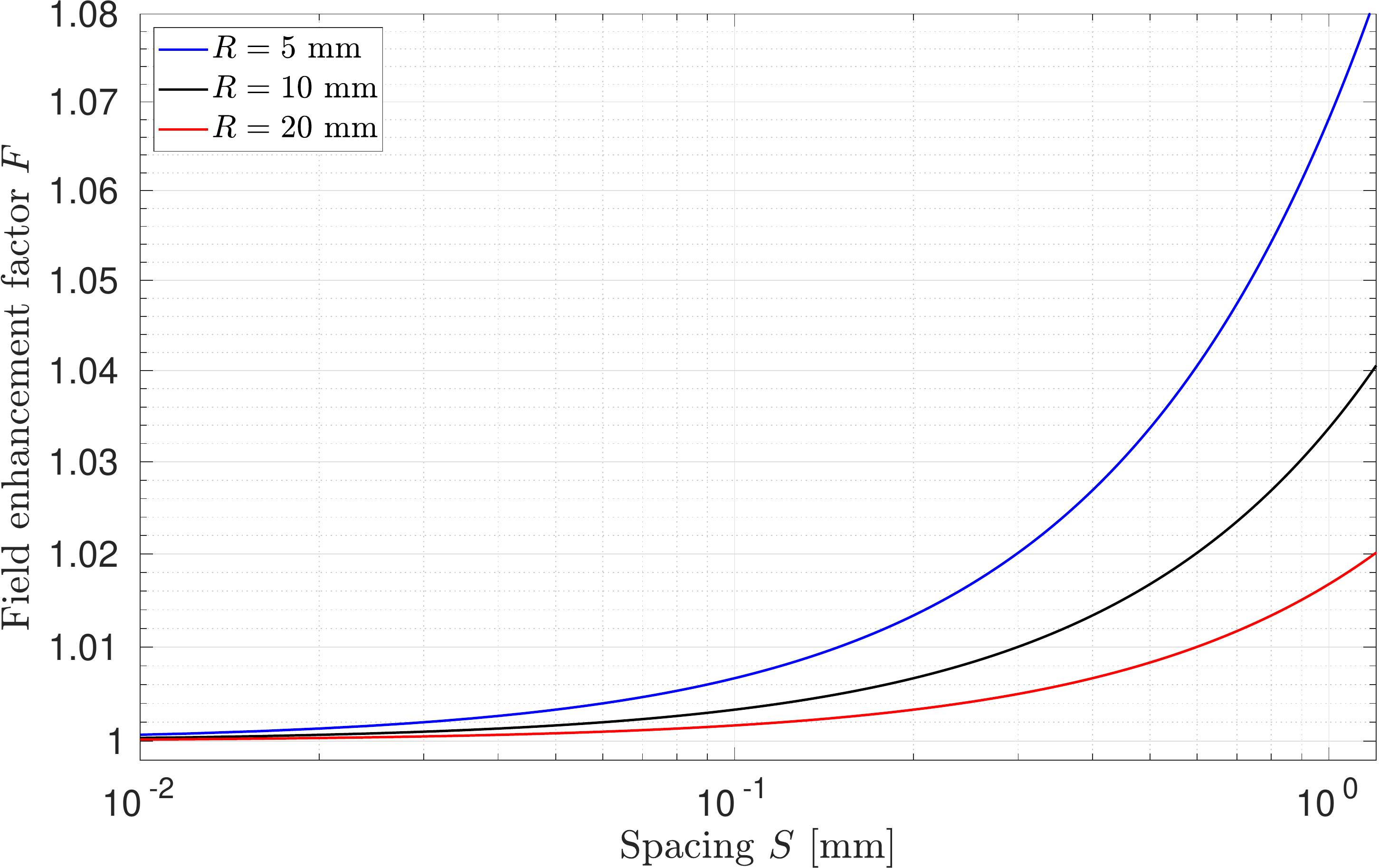}
	\caption{Field enhancement factor $F$ as function of spacing $S$ between two ideal half-spheres of radii $R = 5$, $10$, and \SI{20}{mm}.}
	\label{fig:field-enhancement-factors}
\end{figure}

At the employed distances between $0.1$ and \SI{1}{\milli \meter}, the field enhancement factor $F$ changes only by about 3\% (see Fig.\,\ref{fig:field-enhancement-factors}).
 
\section{Dark current measurements}
\label{sec:measurements}

\begin{figure}[tb]
	\centering
	\includegraphics[width=1\columnwidth]{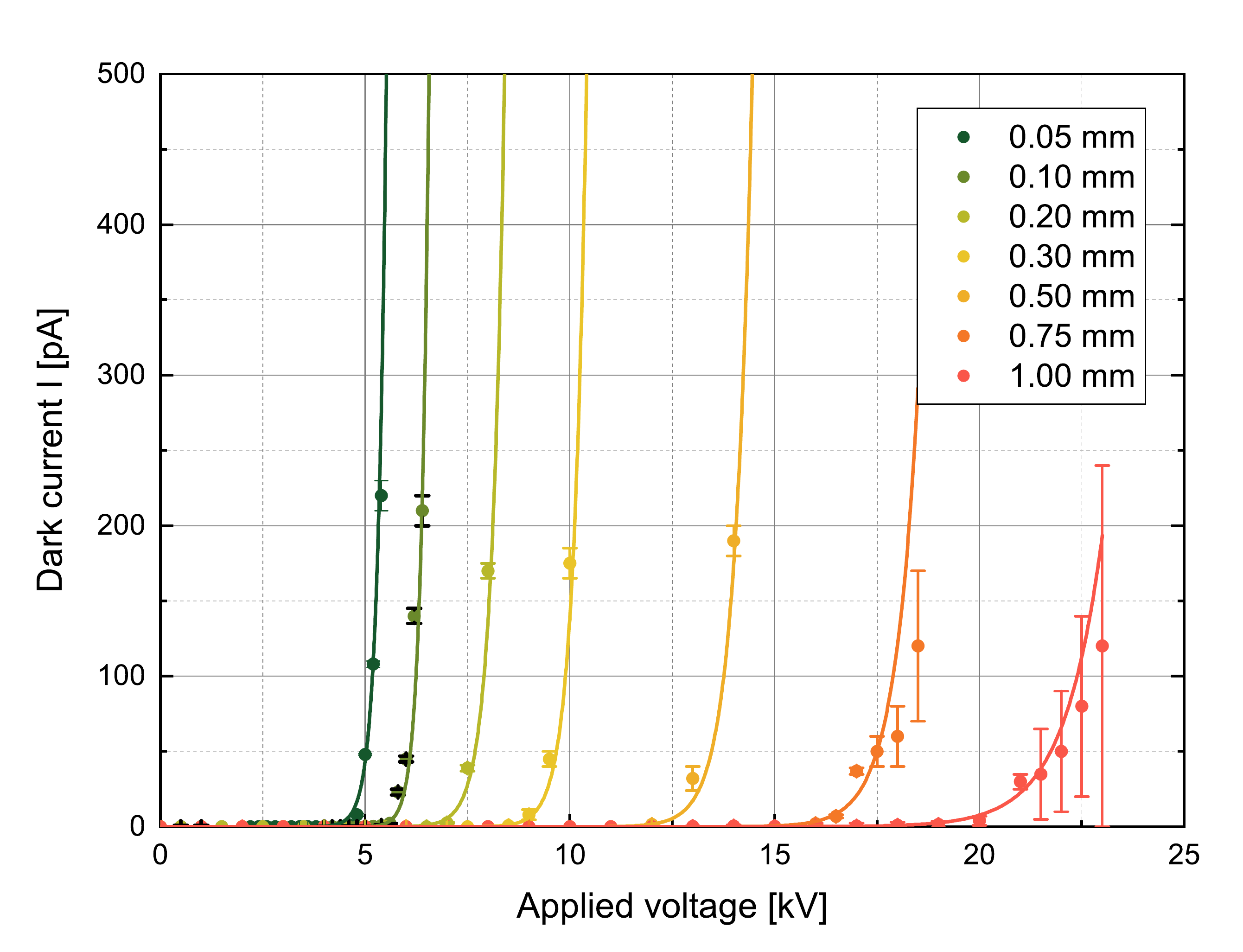}
	\caption{Dark current measured using stainless-steel half-sphere electrodes of $R = \SI{10}{mm}$ radius at distances $S = 0.05$ to \SI{1}{mm}.}
	\label{fig:dark-current-SS}
\end{figure}

For the measurements the experimental setup was transferred to the COSY hall at Forschungszentrum J\"ulich.  The first high-voltage tests were performed with well-polished stainless-steel half-sphere electrodes over a wide range of distances ranging from $S = \SI{30}{mm}$ to $\SI{0.05}{mm}$ (see Fig.\,\ref{fig:dark-current-SS}). Being limited by the \SI{30}{\kilo \volt} power supply, the discharges mainly happened in the test conditions with small gaps between the electrodes. No discharge was observed at a distance $S = \SI{10}{mm}$ with an applied voltage of \SI{30}{\kilo \volt} which leads to $E_\text{max} \approx \SI{4.1}{\mega \volt \per \meter}$. Discharges at larger distances can only be observed when higher voltages are applied which requires a new experimental setup.

For completeness, tests were also carried out by replacing one of the half-sphere electrodes with the larger stainless-steel electrodes with flat surface (see Fig.\,\ref{fig:SS-deflector-prototypes}). In that case, the measured electric field behaved in a similar way and reached values which likely correspond to vacuum breakdown conditions\,\cite{Werner:2004wq,meek1978electrical}.

\begin{figure}[tb]
	\centering
	\includegraphics[width=1\columnwidth]{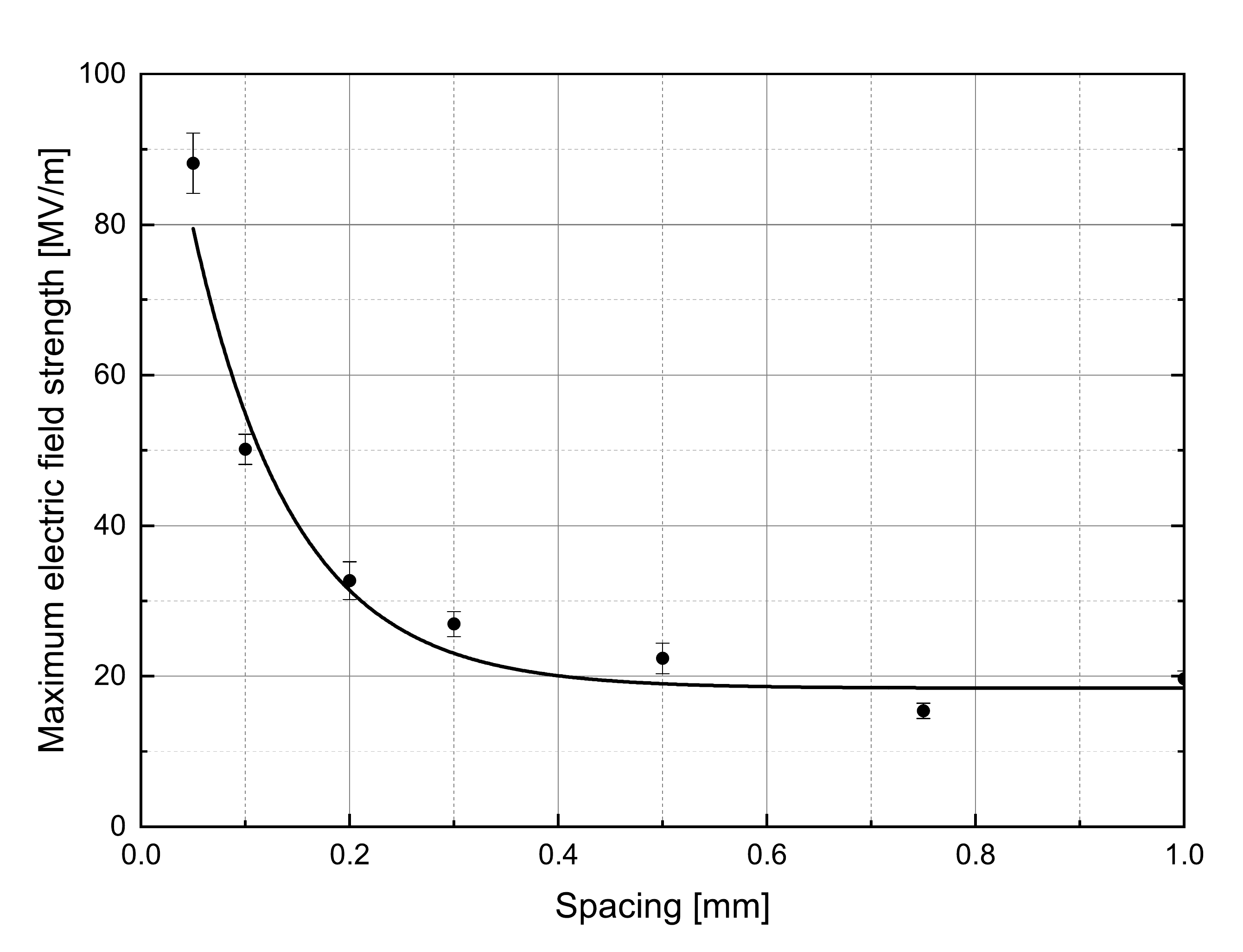}
	\caption{Maximum electric field strength from Fig.\,\ref{fig:dark-current-SS} as function of spacing between the half-sphere stainless-steel electrodes of $R = \SI{10}{mm}$ radius. The values of $E_\text{max}$ computed using Eq.\,(\ref{eq:field-enhancement}) and $F$ from Eq.\,(\ref{eq:field-enhancement-factor}) correspond to zero dark current. The line through the points is drawn to guide the eye.}
	\label{fig:electric-field-strength}
\end{figure}

The measured minimal dark currents were compatible with zero to a tens of a picoampere (see Fig.\,\ref{fig:dark-current-SS}). The maximum values of the electric field $E_\text{max}$, shown in Fig.\,\ref{fig:electric-field-strength} and calculated using Eq.\,(\ref{eq:field-enhancement}) and $F$ from Eq.\,(\ref{eq:field-enhancement-factor}), are taken at the measurement points when the dark current was still compatible with zero within errors. The measurements showed that with half-sphere electrodes of \SI{10}{mm} radius at distances of less than a millimeter, electric fields above the required values of $E = \SI{17}{\mega \volt \per \meter}$ could be reached.

The maximum electric fields obtained at a distance $S = \SI{0.05}{\milli \meter}$, however, are still an order of amplitude smaller than achieved elsewhere\cite{Descoeudres} at much smaller distances of $S = \SI{0.02}{\milli \meter}$. It should be noticed that with respect to the development of electrostatic deflector elements for the future EDM ring, the region of interest ranges from a few cm to about $\SI{10}{\centi \meter}$ distance, which can be studied only with large deflectors and much higher applied voltages.

\section{Summary}
\label{sec:summary}

Mechanically polished stainless-steel electrodes at distances less than a millimeter demonstrate that electric fields close to the breakdown limit in ultra-high vacuum can be reached. The maximum electric fields obtained in the measurements using scaled-down electrodes look promising. They are clearly above the required values for an electrostatic deflector of \SI{17}{\mega \volt \per \meter} for a future EDM ring of \SI{30}{m} radius. The improvement of the HV breakdown capability using different electrodes materials and coatings as well as gas conditioning will be further investigated in the future.

We will now move on to measurements with real-size deflector elements of a length $\ell = \SI{1020}{\milli\meter}$ at distances of $S \approx \SI{20} - \SI{120}{\milli \meter}$ between the plates. A suitable experimental infrastructure with two $\SI{200}{\kilo \volt}$ power supplies is presently set up at IKP of Forschungszentrum J\"ulich.

\begin{acknowledgments}

This work has been performed in the framework of the JEDI collaboration, and is financially supported by an ERC Advanced-Grant (srEDM \# 694390) of the European Union.

\end{acknowledgments}

\bibliographystyle{apsrev4-1}
\bibliography{deflector.bbl}

\end{document}